\documentclass[twocolumn]{aastex631}

\usepackage{amsmath}
\usepackage[mathlines]{lineno}

\begin{document}

\title{The Effect of Age and Stellar Model Choice on Globular Cluster Color-to-Metallicity Conversions}

\author[0000-0001-5290-6275]{Kate Hartman}
\affiliation{Department of Physics \& Astronomy, McMaster University, 1280 Main St W, Hamilton, ON L8S 1T7, Canada}

\author[0000-0001-8762-5772]{William E. Harris}
\affiliation{Department of Physics \& Astronomy, McMaster University, 1280 Main St W, Hamilton, ON L8S 1T7, Canada}

\begin{abstract}
The photometric colors of globular clusters (GCs) act as effective proxies for metallicity, since all normally used optical/IR color indices exhibit a nonlinear but monotonic relation between their integrated color and their metallicity.  One color index, (g - z) or (F475W - F850LP), has been spectroscopically calibrated in several studies, providing leverage to define color-to-metallicity conversions for other indices.  In this paper, building on the work of \citet{hartman2023GCsenviro}, we study the GC color-metallicity relation in more detail by testing the dependence of the relations on different suites of stellar models and different assumed GC ages.  Though noticeable differences between models exist, we find that the net effect on the derived GCS metallicity distributions is small.
\end{abstract}

\keywords{Globular star clusters (656) --- Metallicity (1031) --- Stellar colors (1590)}

\section{Introduction} \label{sec:intro}

Globular clusters (GCs)---old, massive, dense star clusters found in all but the least massive of dwarf galaxies \citep{harris2010review,forbes2018metgrad,beasley2020review,eadie2022dwarfgcs}---are powerful tracers of early galaxy growth and chemical enrichment.  Part of their appeal for observers stems from the scaling relations that they exhibit with galaxy properties, such as the $M_{GCS}$-$M_h$ relation \citep[e.g.][]{blakeslee1997scaling,forbes2016scaling,harris2017BCGs,dornan2023scaling}{}{}, and from the relations between fundamental GC properties and easy-to-observe quantities.  The relation between GC metallicity and color, for example, is monotonic \citep[see][among others]{brodie2006colour,peng2006nonlinear,usher2015nonlinear,harris2017BCGs,fahrion2020spectroscopy,harris2023colours}, making color an attractive proxy observable for metallicity.

Despite its usefulness, the specifics of the GC color-metallicity relation (CMR) have proven challenging to constrain.  When spectroscopy is available, the relation is often modelled nonlinearly, as in \citet{peng2006nonlinear,sinnott2010colour,usher2012colour,fahrion2020spectroscopy}, and \citet{harris2023colours}  \citep[with the exception of][]{villaume2019spectroscopy}.  Furthermore, the relation has been spectroscopically studied primarily for the (g - z) color index (equivalent to the HST index (F475W - F850LP) in the Vegamag system); thus other color indices are useful metallicity indicators but must first be transformed to (g - z) with a color-color relation, as in \citet{harris2023colours} and \citet{hartman2023GCsenviro}.

Both \citet{harris2023colours} and \citet{hartman2023GCsenviro} used the PARSEC stellar models \citep{marigo2017parsec} to derive GC color-color relations.  They adopted a fixed age of 12 Gyr for their simulated clusters and fitted simple nonlinear models to the resulting CMRs.  Those studies focused on the GC metallicity distributions as they related to host galaxy properties; while the steps taken to construct color-color and color-metallicity relations were justifiable, the relations were a means to an end and were not themselves investigated in detail.  Because GC integrated colors are dominated by relatively small numbers of red-giant and subgiant stars per cluster, stellar model specifics in these evolutionary stages could have a significant impact on the shape of GC color-color relations, with subsequent effects carried through the CMRs.

In this paper, we take further steps towards building the CMRs for old GCs by investigating the effect of stellar model choices on the shape and zeropoints of the relations.  For the purposes of this study, we restrict the analysis to the particular color index (F475X-F110W) used in \citet{hartman2023GCsenviro} for the GC systems in 15 giant early-type galaxies (ETGs), with our main goal to test the model-based transformations themselves.  We will address extensions to other indices in later work.  Section \ref{sec:modelGCs} details the properties of our simulated clusters, Section \ref{sec:colourmet} outlines our procedures for deriving CMRs, and Section \ref{sec:observations} looks at how the differences in CMRs affect analysis of observations.  We discuss our findings and summarize our work in Section \ref{sec:conclusions}.

\section{Model GCs}
\label{sec:modelGCs}

We derived GC CMRs using five suites of frequently used contemporary models: two versions of the PARSEC (PAdova and TRieste Stellar Evolution Code) models, two versions of the BaSTI (a Bag of Stellar Tracks and Isochrones) models, and the MIST (MESA Isochrones and Stellar Tracks) models.

First, we tested two recent versions of the PARSEC cluster simulator: CMD 3.6, the model used in \citet{hartman2023GCsenviro}, and CMD 3.7, the most up-to-date set available at the time of writing.  We used the default settings from the Osservatorio Astronomico di Padova's online tool (\href{http://stev.oapd.inaf.it/cgi-bin/cmd_3.6}{CMD 3.6}, \href{http://stev.oapd.inaf.it/cgi-bin/cmd_3.7}{CMD 3.7}) for circumstellar dust, extinction, and long period variability, and specified the cluster age, metallicity, and mass.  See \citet{bressan2012parsec,chen2014parsec,chen2015parsec,tang2014parsec,marigo2017parsec,pastorelli2019parsec}, and \citet{pastorelli2020parsec} for details.

Next, we tested two abundance mixtures from BaSTI: solar-scaled, with [$\alpha$/Fe] = 0.0, and alpha-enhanced, with [$\alpha$/Fe] = +0.4, both available as drop-down options on the synthetic CMD page of BaSTI's \href{http://basti-iac.oa-teramo.inaf.it/syncmd.html}{online tool}.  We used the diffusionless grid for our solar-scaled clusters and the He = 0.247 grid for our alpha-enhanced clusters, along with default values for star formation rate (SFR), minimum mass, binary fraction, minimum binary mass ratio, treatment of variable stars, and photometric error.  We also specified an SFR scale of 500,000 (this input is related to the mass of the resulting simulated cluster).  BaSTI is the only model suite in this study with an alpha enhancement option as of the time of writing.  See \citet{hidalgo2018basti,pietrinferni2021basti}, and \citet{salaris2022basti} for details.

Finally, we tested one set of MIST models with MIST's isochrone interpolation \href{https://waps.cfa.harvard.edu/MIST/interp_isos.html}{online tool}.  We used the default settings for MIST version, stellar rotation, extinction, and abundances.  See \citet{dotter2016mist} and \citet{choi2016mist}, along with \citet{paxton2011mesa,paxton2013mesa,paxton2015mesa}, and \citet{paxton2018mesa} for details.

For each of the 5 suites of stellar models, we created a $5 \times 23$ grid of simulated clusters as they would appear in the HST WFC3 photometric system, with a metallicity range of -2.2 to 0.0 dex in 0.1-dex steps and an age range of 9 to 13 Gyr in 1-Gyr steps.  Each simulated cluster had approximately the same mass ($M \sim 10^5 M_{\odot}$), a Kroupa initial mass function, and no internal spread in metallicity [M/H].  We used these simulated cluster sets to calculate predicted CMRs.

\section{Deriving CMRs}
\label{sec:colourmet}

In order to recreate the color to metallicity conversion construction process from \citet{hartman2023GCsenviro}, we needed integrated magnitudes from simulated clusters for each of the filters in their color indices.  The BaSTI online tool provides these automatically, producing two outputs for each model cluster: a large file containing magnitudes and stellar properties for each simulated star, and a small file with cluster properties and integrated magnitudes.  Conversely, the MIST online tool produces isochrones rather than simulated clusters, and the PARSEC CMD 3.6 and 3.7 online tools' integrated magnitude option removes the cluster mass parameter, so we calculated integrated magnitudes for each PARSEC and MIST cluster manually.

\subsection{Integrated cluster magnitudes}

The output files for both PARSEC CMD 3.6 and 3.7 included magnitudes in all HST WFC3 filters for each individual star in the simulated cluster.  We converted these magnitudes to luminosities, summed them to get an integrated cluster luminosity, and then converted that back into a magnitude for each relevant filter.

For MIST, an output file contains for each point on the isochrone a set of theoretical stellar parameters (e.g. mass, effective temperature, luminosity, etc.).  For the MIST models, we calculated the number of stars at each point necessary for a $10^5 M_{\odot}$ GC with a Kroupa IMF \citep{kroupa1993imf}, and then followed the same luminosity summing procedures that we used with the PARSEC clusters.

\begin{figure*}
    \centering
    \includegraphics[width=\textwidth]{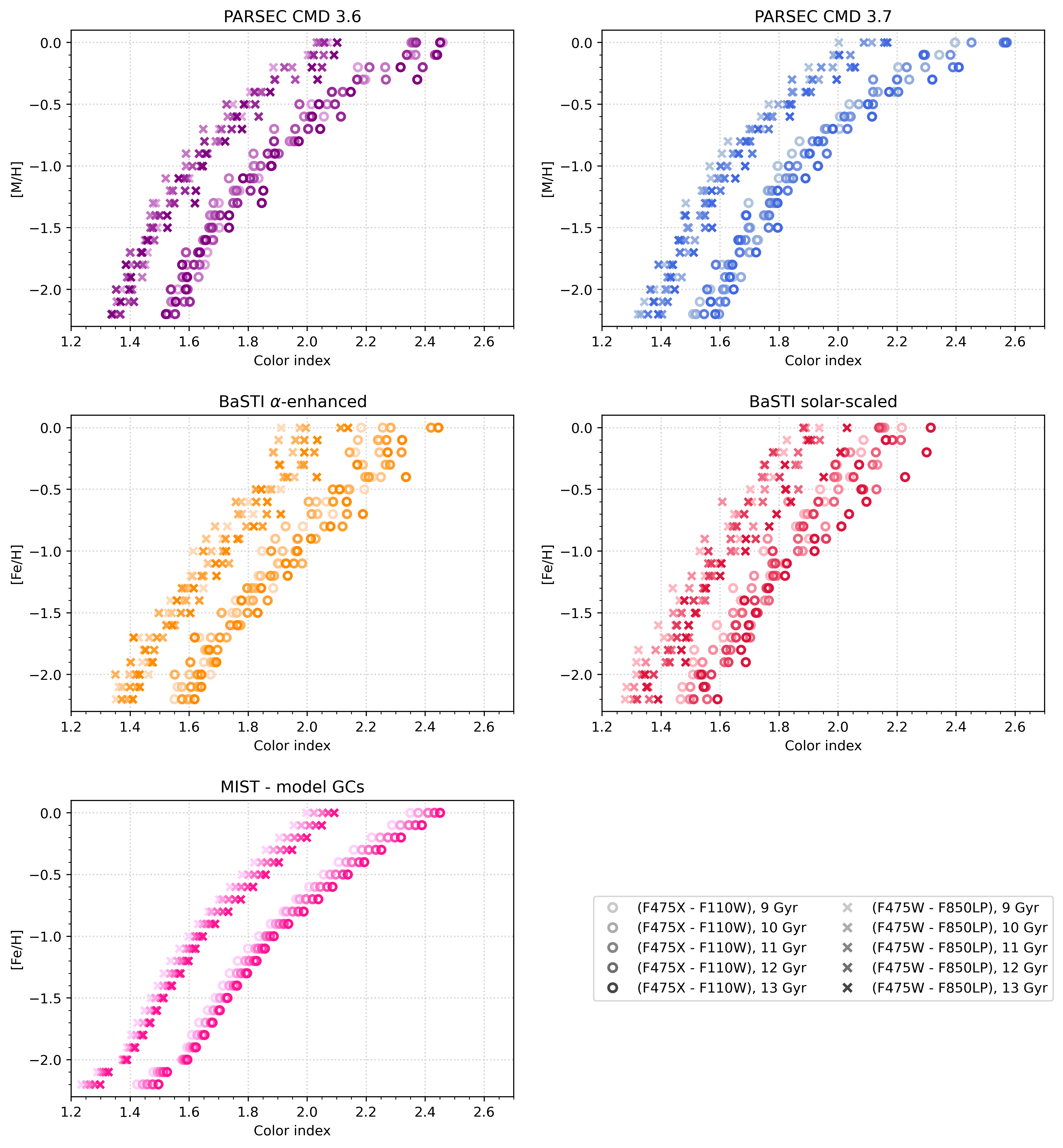}
    \caption{Model GCs from the PARSEC, BaSTI, and MIST SSPs.  With all five models, we simulated GCs with metallicities ranging from -2.2 to 0.0 dex and ages ranging from 9 to 13 Gyr, and calculated integrated magnitudes in (F475X - F110W) and (F475W - F850LP).  Top left: PARSEC CMD 3.6, the model used in \citet{hartman2023GCsenviro}.  Top right: PARSEC CMD 3.7, the latest version of PARSEC available online as of the time of writing.  Middle left: BaSTI's $\alpha$-enhanced model.  Middle right: BaSTI's solar-scaled model.  Bottom left: MIST isochrone-derived model GCs.}
    \label{fig:modelGCs}
\end{figure*}

Figure \ref{fig:modelGCs} shows (F475X - F110W) and (F475W - F850LP), the two color indices from \citet{hartman2023GCsenviro}, versus metallicity for individual simulated GCs in each of the five models.  Because the MIST online tool does not produce simulated clusters like its PARSEC and BaSTI counterparts do, those clusters are not affected by stochasticity and do not exhibit any scatter.

\subsection{Fitting}

\begin{figure}
    \centering
    \includegraphics[width=0.47\textwidth]{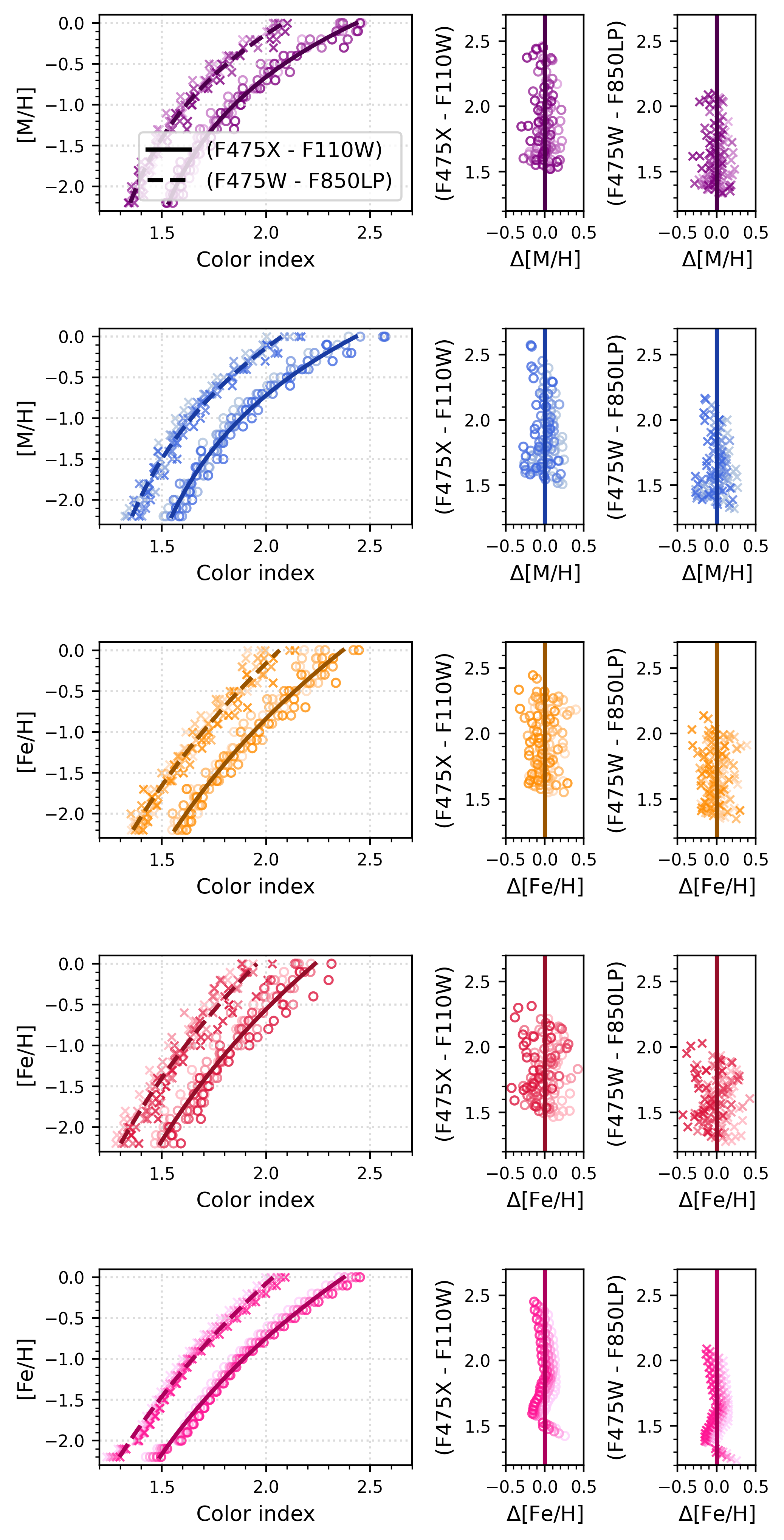}
    \caption{\textit{Left column:} model GCs with fitted exponential metallicity-color relations overplotted.  Points are colored as in Figure \ref{fig:modelGCs}.  \textit{Center column:} residuals from the (F475X - F110W)-metallicity relations.  \textit{Right column:} residuals from the (F475W - F850LP)-metallicity relations.}
    \label{fig:metcolfitsoverplot}
\end{figure}

\begin{figure*}
    \centering
    \includegraphics[width=\textwidth]{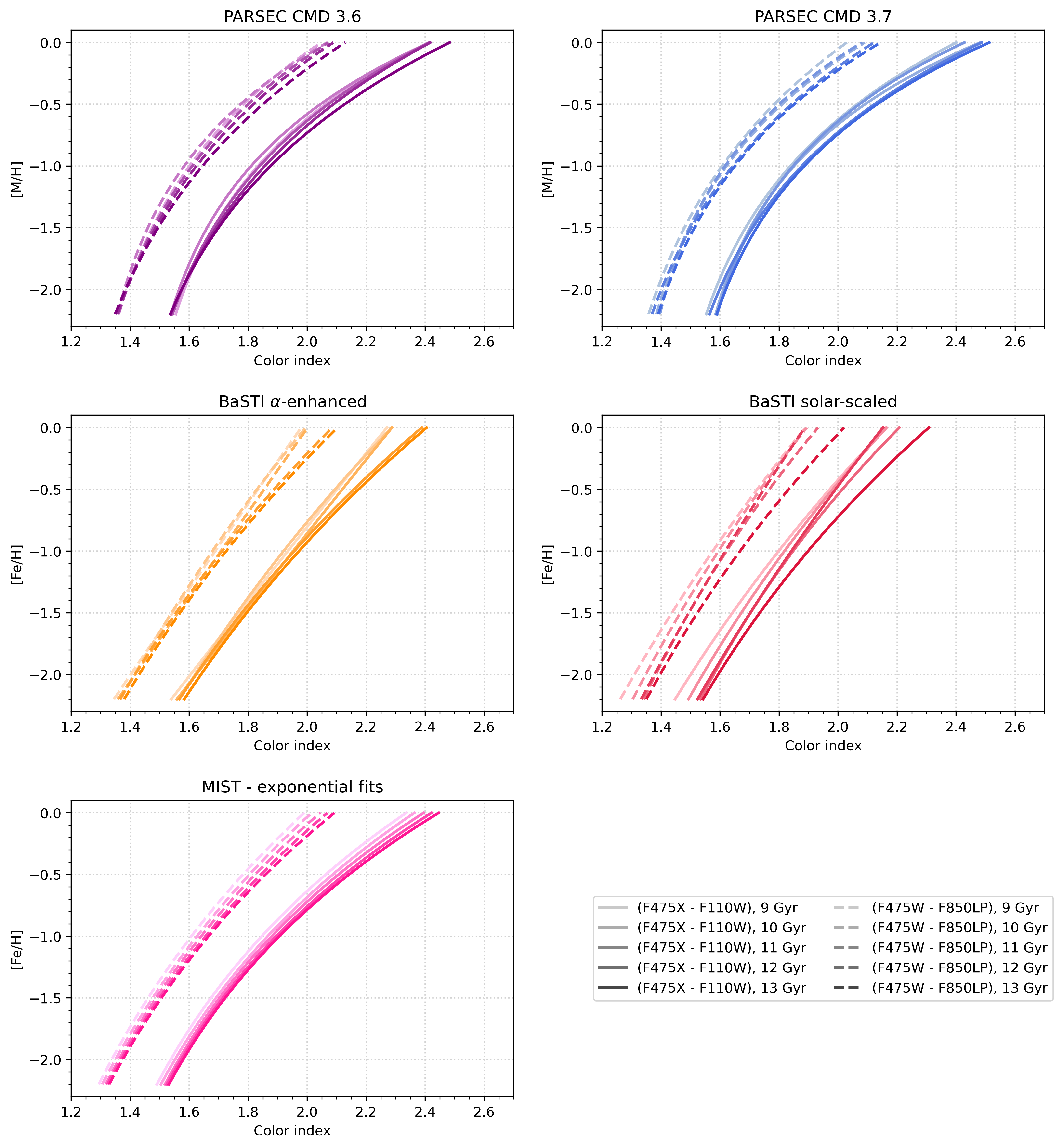}
    \caption{Fitted exponential metallicity-color relations for the five models and the five GC age groups.}
    \label{fig:metcolfits}
\end{figure*}

\begin{figure}
    \centering
    \includegraphics[width=0.47\textwidth]{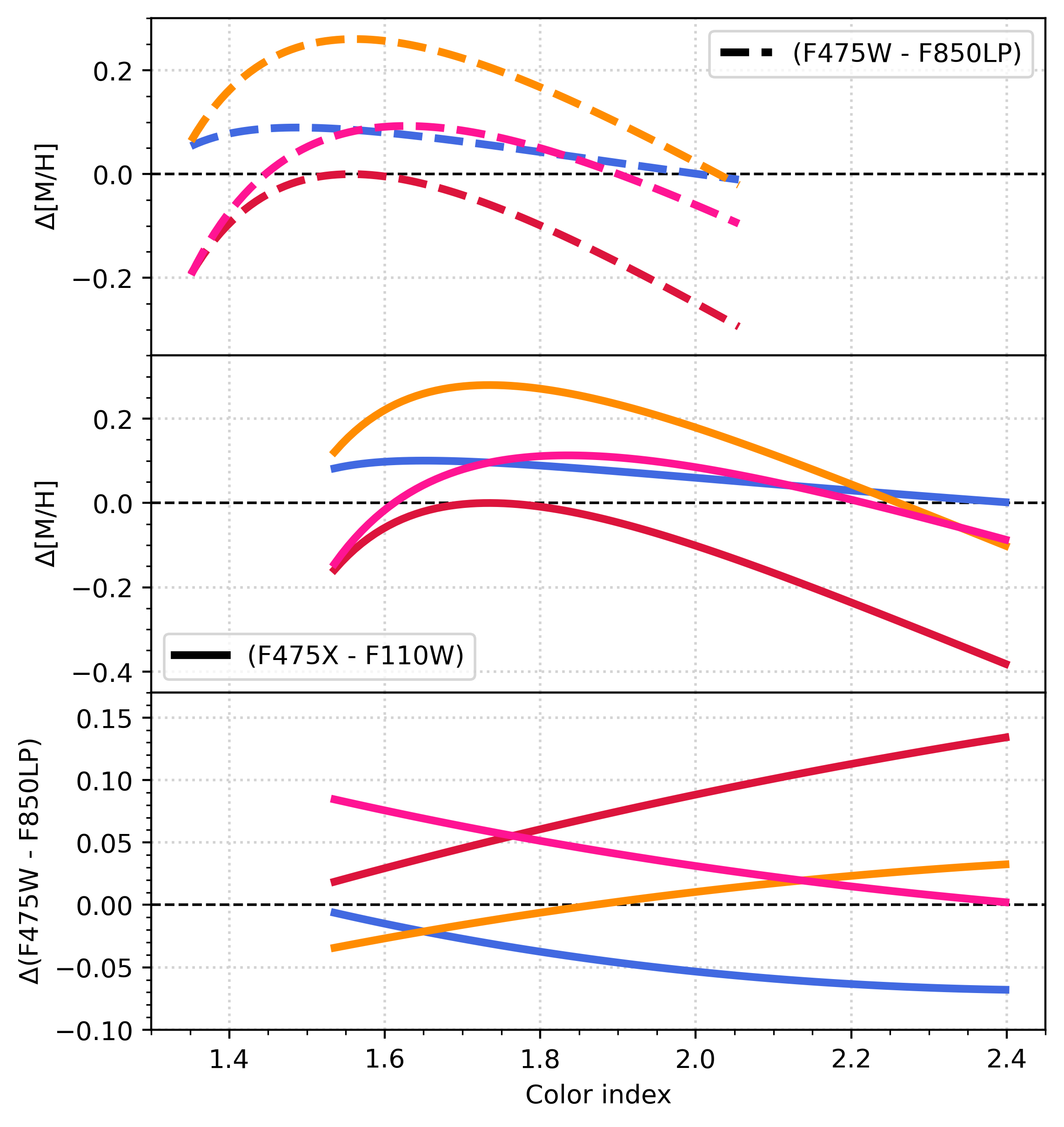}
    \caption{Differences between fitted exponential color-metallicity and color-color relations, comparing PARSEC CMD 3.6 to the other four models (i.e. subtracting each model from PARSEC CMD 3.6).  As in previous figures, PARSEC CMD 3.7 is shown in blue, BaSTI alpha-enhanced in orange, BaSTI solar-scaled in red, and MIST in pink; similarly, relations based on the (F475X - F110W) color index are shown as solid lines and relations based on the (F475W - F850LP) index as dashed lines.}
    \label{fig:colcolfits}
\end{figure}

We fitted inverse exponential equations to each set of simulated GCs in both (F475X - F110W) and (F475W - F850LP), for a total of 10 fitted equations per stellar model with 5 per color index.  We also fitted all simulated GCs regardless of age from each stellar model; the fitted parameters are defined in Equation \ref{eq:invexp} and listed in Table \ref{tab:nominal_fits}.  An exponential model is preferable because unlike a quadratic model it remains monotonic even when used to extrapolate (although significant extrapolations should still be viewed with caution), and unlike a pair of joined linear models it assumes no sudden change in slope.  Figure \ref{fig:metcolfitsoverplot} shows model GCs of all ages with fitted exponential models overplotted, along with residuals for the CMRs for both color indices; Figure \ref{fig:metcolfits} places all fitted models in metallicity-color space; and Figure \ref{fig:colcolfits} shows differences when subtracting the color-metallicity and color-color relations of the other four models from the PARSEC CMD 3.6 color-metallicity and color-color relations that we use as a template.  (That is, for a given value in (F475X - F110W), how much more metal-poor or metal rich, or bluer or redder in (F475W - F850LP), is the alternate model compared to PARSEC CMD 3.6?)  The synthetic color-color relations show agreement within 0.15 magnitudes among the different sets of models, and the CMRs agree within 0.3 dex.

The full set of fitted parameters, broken out by GC age, for an equation of the form
\begin{equation}
    \label{eq:invexp}
    \text{[M/H]} = \frac{\ln{\frac{\text{color} - C}{A}}}{B}
\end{equation} can be found in Appendix \ref{app:relation_params}.

\subsection{CMR uncertainty}

\begin{figure}
    \centering
    \includegraphics[width=0.47\textwidth]{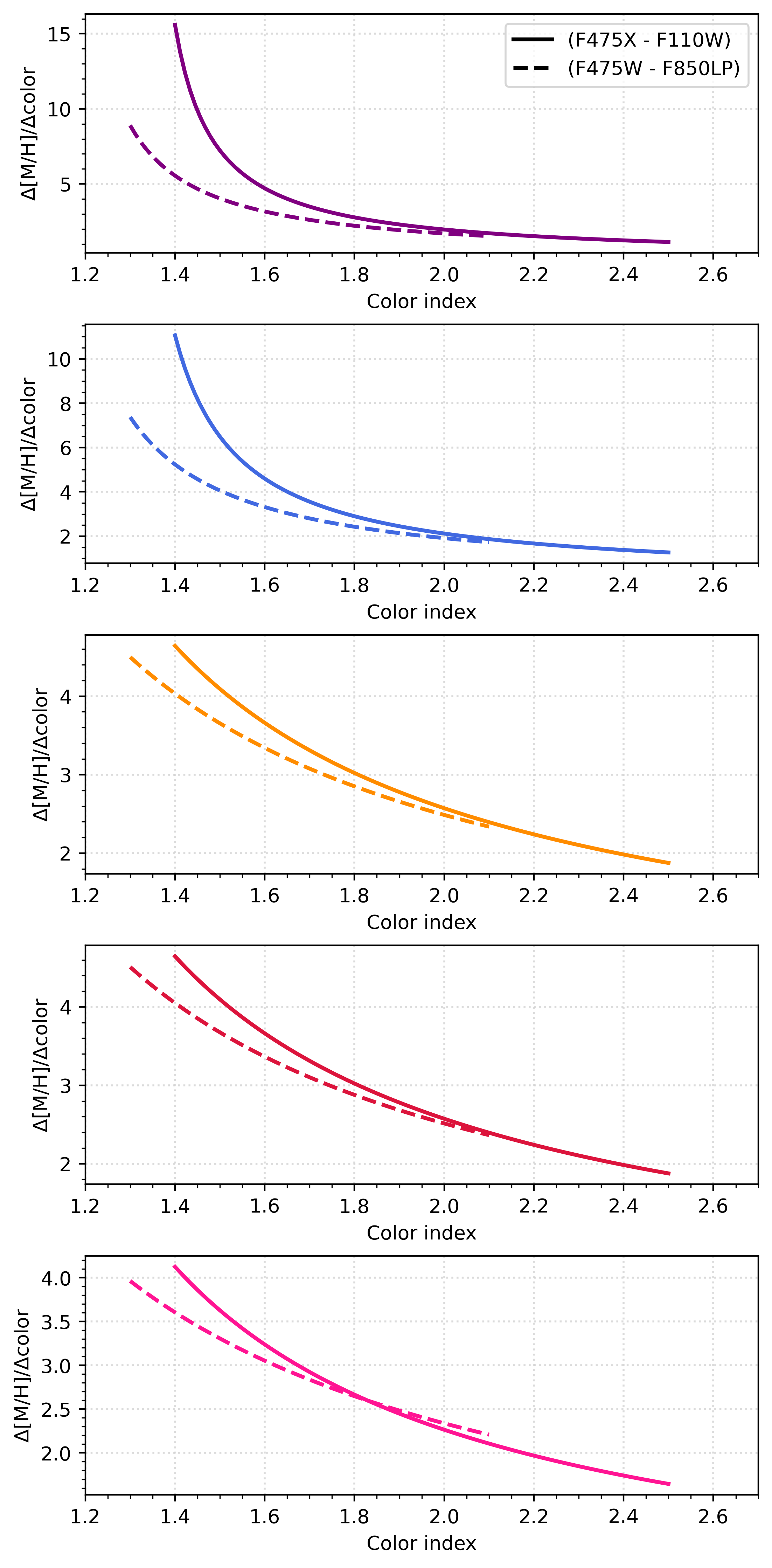}
    \caption{Ratio of metallicity uncertainty and color uncertainty by color index for each of the five stellar models and both color indices; see Equation \ref{eq:deltaMH}.  Color is a less sensitive tracer of metallicity at the blue end of color-metallicity space than at the red end.}
    \label{fig:dmdc}
\end{figure}

The uncertainties of metallicity values derived from the CMRs presented in this work can be estimated by multiplying a user-supplied color uncertainty $\Delta \text{color}$ by the slope of the CMR $\frac{d \text{[M/H]}}{d \text{color}}$:
\begin{equation}
    \label{eq:deltaMH}
    \Delta \text{[M/H]} = \Delta \text{color} \frac{d \text{[M/H]}}{d \text{color}}
\end{equation} where
\begin{equation}
    \label{eq:dmdc}
    \frac{d \text{[M/H]}}{d \text{color}} = \frac{1}{B(\text{color} - C)}
\end{equation} with B and C found in Table \ref{tab:nominal_fits} as with Equation \ref{eq:invexp}.  Figure \ref{fig:dmdc} shows the ratio of metallicity uncertainty to color uncertainty for each model in (F475X - F110W) and (F475W - F850LP); 
because of the change in slope of the CMRs, 
the red ends of the CMRs are more sensitive to metallicity than the blue ends, so the metallicity uncertainties at the red ends are smaller.

\begin{table*}
	\centering
	\caption{CMR parameters (as defined in Equation \ref{eq:invexp}) for each stellar model and both color indices, derived from all five simulated GC age groups.}
	\label{tab:nominal_fits}
    \scriptsize
	\begin{tabular}{ll|ccc}
		\hline
		Index & Model & A & B & C\\
        \hline
		(F475X - F110W) & PARSEC CMD 3.6 & 1.120 ($\pm$ 0.024) & 0.741 ($\pm$ 0.051) & 1.313 ($\pm$ 0.031)\\
		  & PARSEC CMD 3.7 & 1.173 ($\pm$ 0.037) & 0.636 ($\pm$ 0.053) & 1.258 ($\pm$ 0.045)\\
		  & BaSTI $\alpha$-enhanced & 1.715 ($\pm$ 0.293) & 0.289 ($\pm$ 0.072) & 0.654 ($\pm$ 0.306)\\
          & BaSTI solar-scaled & 1.581 ($\pm$ 0.307) & 0.289 ($\pm$ 0.082) & 0.655 ($\pm$ 0.320)\\
          & MIST & 1.703 ($\pm$ 0.117) & 0.332 ($\pm$ 0.035) & 0.670 ($\pm$ 0.124)\\
		\hline
        (F475W - F850LP) & PARSEC CMD 3.6 & 0.950 ($\pm$ 0.025) & 0.678 ($\pm$ 0.052) & 1.134 ($\pm$ 0.032)\\
		  & PARSEC CMD 3.7 & 1.020 ($\pm$ 0.044) & 0.554 ($\pm$ 0.054) & 1.055 ($\pm$ 0.052)\\
		  & BaSTI $\alpha$-enhanced & 1.632 ($\pm$ 0.319) & 0.257 ($\pm$ 0.070) & 0.434 ($\pm$ 0.330)\\
          & BaSTI solar-scaled & 1.538 ($\pm$ 0.364) & 0.251 ($\pm$ 0.082) & 0.418 ($\pm$ 0.376)\\
          & MIST & 1.740 ($\pm$ 0.190) & 0.251 ($\pm$ 0.038) & 0.293 ($\pm$ 0.196)\\
        \hline
	\end{tabular}
\end{table*}

\subsection{The GC age-metallicity relation}

GC metallicity is expected on both theoretical and observational grounds to be correlated at least slightly with age, in the sense that more metal-rich clusters, on average, belong to a later stage of formation and are thus younger \citep{leaman2013agemet,choksi2018agemet,li2020agemet,horta2021agemet}.
To model the correlation between GC metallicity and age, we added an age-based correction to our color-metallicity procedure.

We calibrated the GC age-metallicity relation by combining data for Milky Way GCs from \citet{dotter2010MWGCs}, \citet{forbes2010MWGCs}, and \citet{vandenberg2013MWGCs}, and finding a linear fit to the aggregated dataset:

\begin{equation}
\label{eq:agefit}
    \text{age (Gyr)} = -0.744\text{[M/H]} + 11.38
\end{equation}

\begin{figure}
    \centering
    \includegraphics[width=0.47\textwidth]{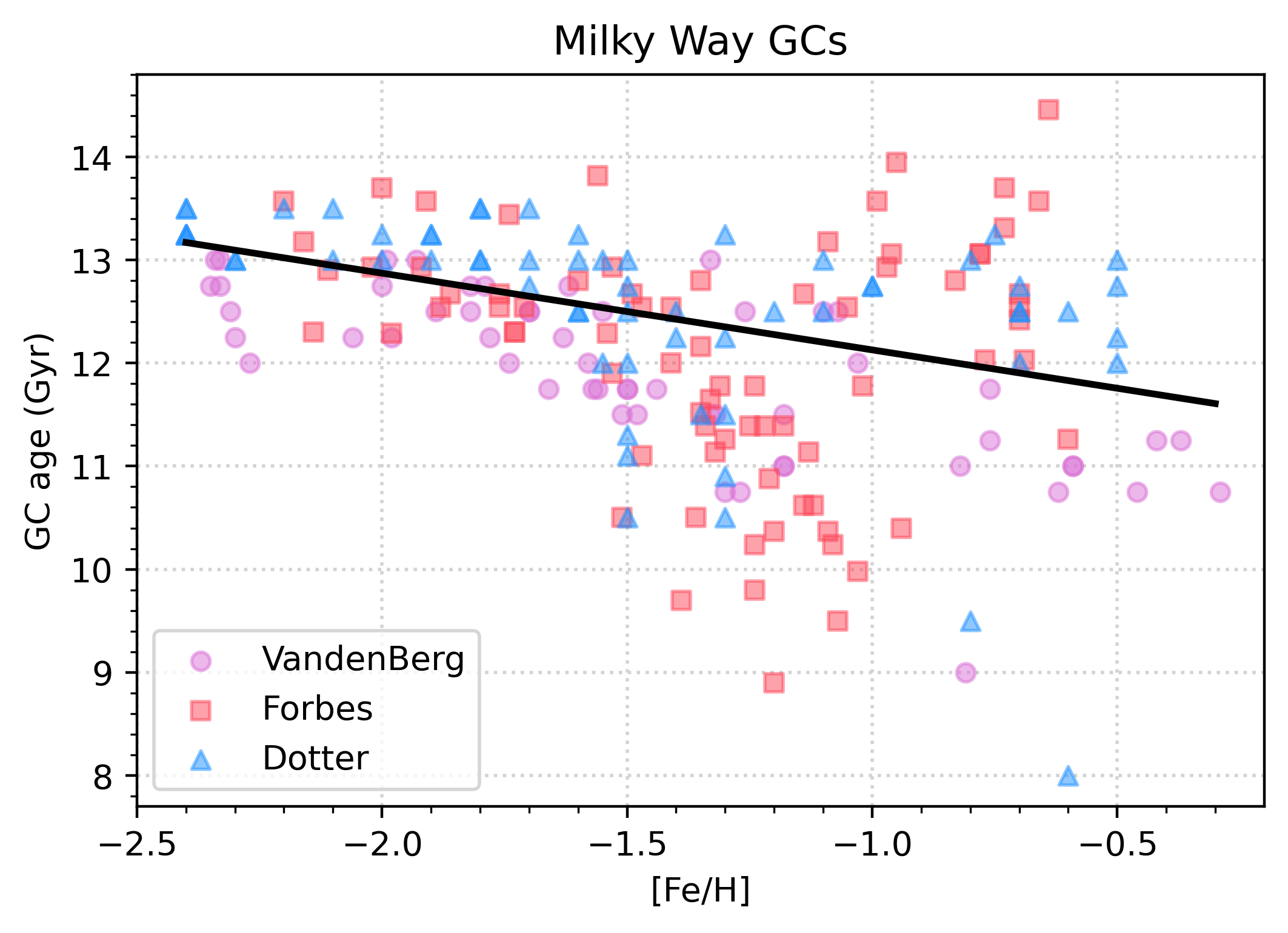}
    \caption{Age vs. metallicity for Milky Way GCs.  Data are from \citet{vandenberg2013MWGCs}, \citet{forbes2010MWGCs}, and \citet{dotter2010MWGCs}; the black line shows Equation \ref{eq:agefit}, a fitted linear age-metallicity relation based on all three datasets.}
    \label{fig:MW_agemet}
\end{figure}

Figure \ref{fig:MW_agemet} shows the original data with Equation \ref{eq:agefit} overplotted in black.  Though there is no guarantee that GCs in other large galaxies, such as the giant early-type galaxies studied in \citet{harris2023colours} and \citet{hartman2023GCsenviro}, will follow this same mean relation, its general property that metal-richer GCs should be younger on average is expected to hold (cf. the references cited above).  The main purpose of the present study is, instead, to explore the sensitivity of the resulting GC metallicity distribution function to an age-metallicity relation.

Equation \ref{eq:agefit} estimates a GC's age based on its metallicity.  To determine the corresponding adjustment from the nominal CMRs listed in Table \ref{tab:nominal_fits} for each stellar model, we divided the average color index range by our model age range (i.e. 4 Gyr) to determine an age-based color offset.  After finding an individual GC's color offset based on its age and adding it to the measured GC color, we applied the nominal CMR again and obtained an age-corrected metallicity value.

\begin{figure*}
    \centering
    \includegraphics[width=\textwidth]{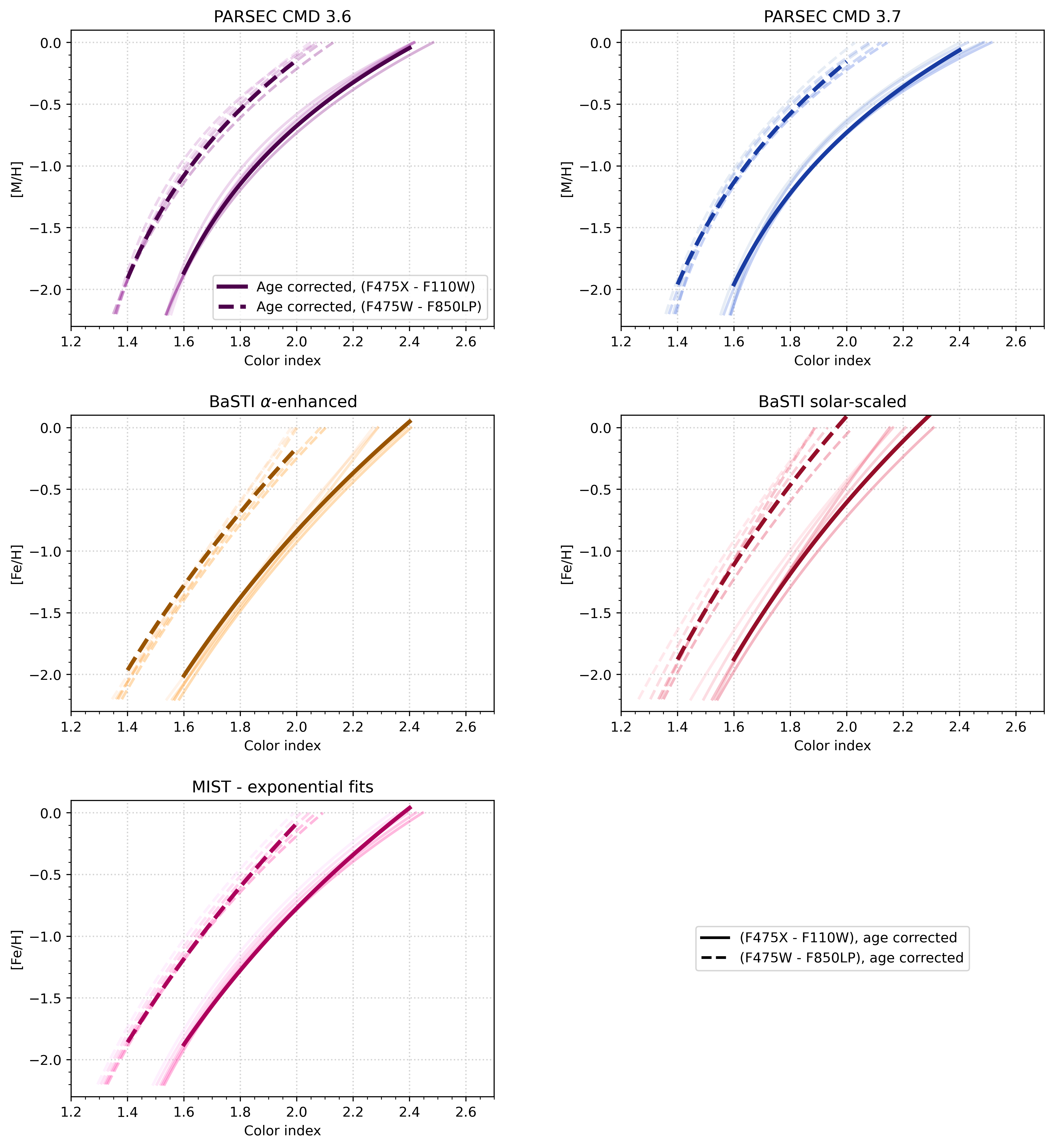}
    \caption{GC age corrections for the fitted CMRs.  The corrected relations are slightly flatter than the CMRs for specific GC ages in the PARSEC and MIST models, and slightly more curved in the BaSTI models.  In all cases, the difference between fitted relations for individual ages and the age-corrected relation is very small.}
    \label{fig:agecorr}
\end{figure*}

Across all five stellar models, the age correction made very little difference to the CMR (see Figure \ref{fig:agecorr}): all corrections were less than 0.05 magnitudes.  The age-corrected relations were slightly flatter than the original relations, with the extreme red and blue ends of the relation moving to less extreme metallicity values, but never leaving the color bounds set by the oldest and youngest model clusters.

\section{Application to observations and discussion}
\label{sec:observations}

In order to compare the performance of each of our CMRs, we tested them on the data from \citet{hartman2023GCsenviro}.  Their sample comprised 15 massive elliptical galaxies with densely populated GCSs, at distances ranging from approximately 60 to 110 Mpc.  Because GCs appear as unresolved point sources to \textit{HST}'s WFC3 camera at those distances, they extracted photometry using DOLPHOT \citep{dolphin2000dolphot} and calculated integrated (F475X - F110W) color indices for each GC in their sample.  After cleaning their sample and adjusting for completeness \citep[see Section 3 of][for details]{hartman2023GCsenviro}, they transformed their color indices into metallicity values for comparison to other variables of interest.  The original images are available at MAST: \dataset[10.17909/pvve-1002]{\doi{10.17909/pvve-1002}}.

\begin{figure*}
    \centering   \includegraphics[width=\textwidth]{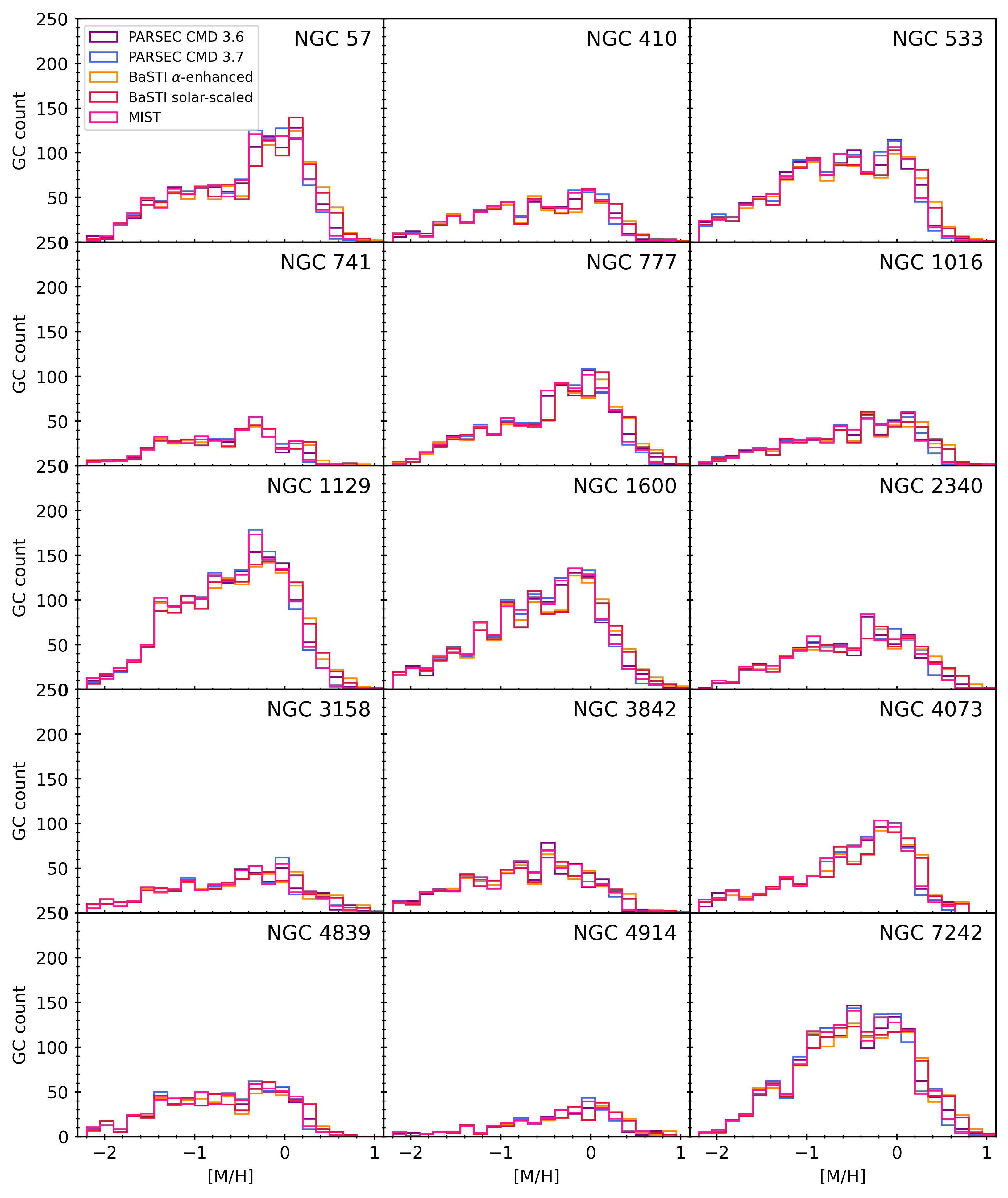}
    \caption{MDFs for the sample from \citet{hartman2023GCsenviro}, transformed from color distribution functions using each of the five stellar models.}
    \label{fig:methists}
\end{figure*}

\begin{figure*}
    \centering   \includegraphics[width=\textwidth]{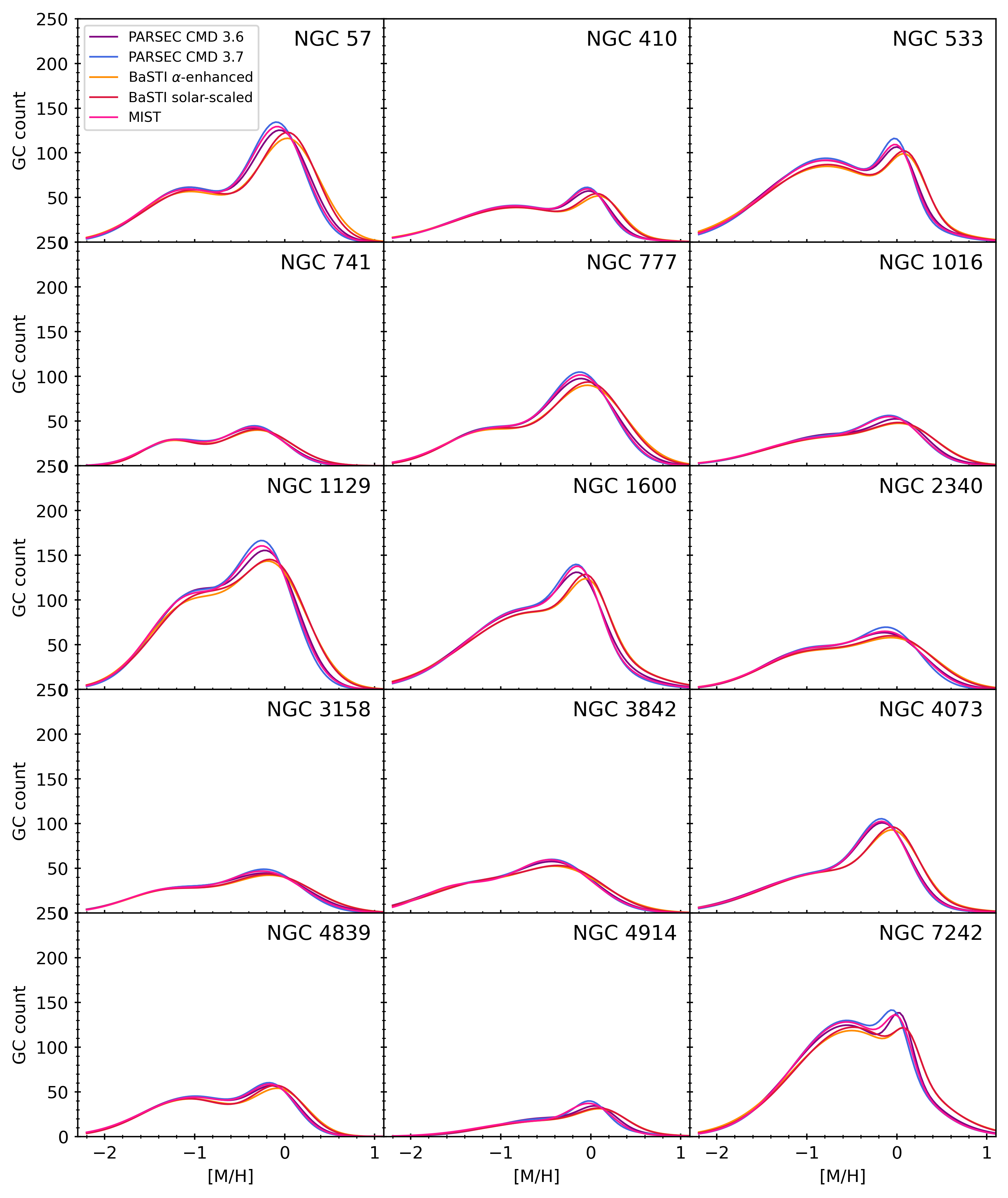}
    \caption{Double Gaussian fits for the MDFs from Figure \ref{fig:methists}.}
    \label{fig:DGfits}
\end{figure*}

We followed the same analysis procedure from that work, but substituted in our alternate CMRs before using GMM \citep{muratov2010gmm} to fit double Gaussian curves:

\begin{equation}
    N_{GCs,comp} = A_{mp}e^{{\frac{-(x-\mu_{mp})^2}{2\sigma_{mp}^2}}} + A_{mr}e^{{\frac{-(x-\mu_{mr})^2}{2\sigma_{mr}^2}}}
    \label{eq:DG}
\end{equation} where $A_{mp}$ and $A_{mr}$ are amplitudes, $\mu_{mp}$ and $\mu_{mr}$ are peak positions, and $\sigma_{mp}$ and $\sigma_{mr}$ are peak widths for the metal-poor and metal-rich Gaussian curves, respectively.  Figure \ref{fig:methists} shows metallicity histograms for each of the 15 galaxies in their sample, and Figure \ref{fig:DGfits} shows double Gaussian fits to each metallicity distribution function (MDF) \citep[standard for GCSs; see e.g.][]{kim2013bimodal,cantiello2014bimodal,brodie2014bimodal,fensch2014bimodal,escudero2015bimodal,harris2017BCGs}.

\begin{figure*}
    \centering   \includegraphics[width=\textwidth]{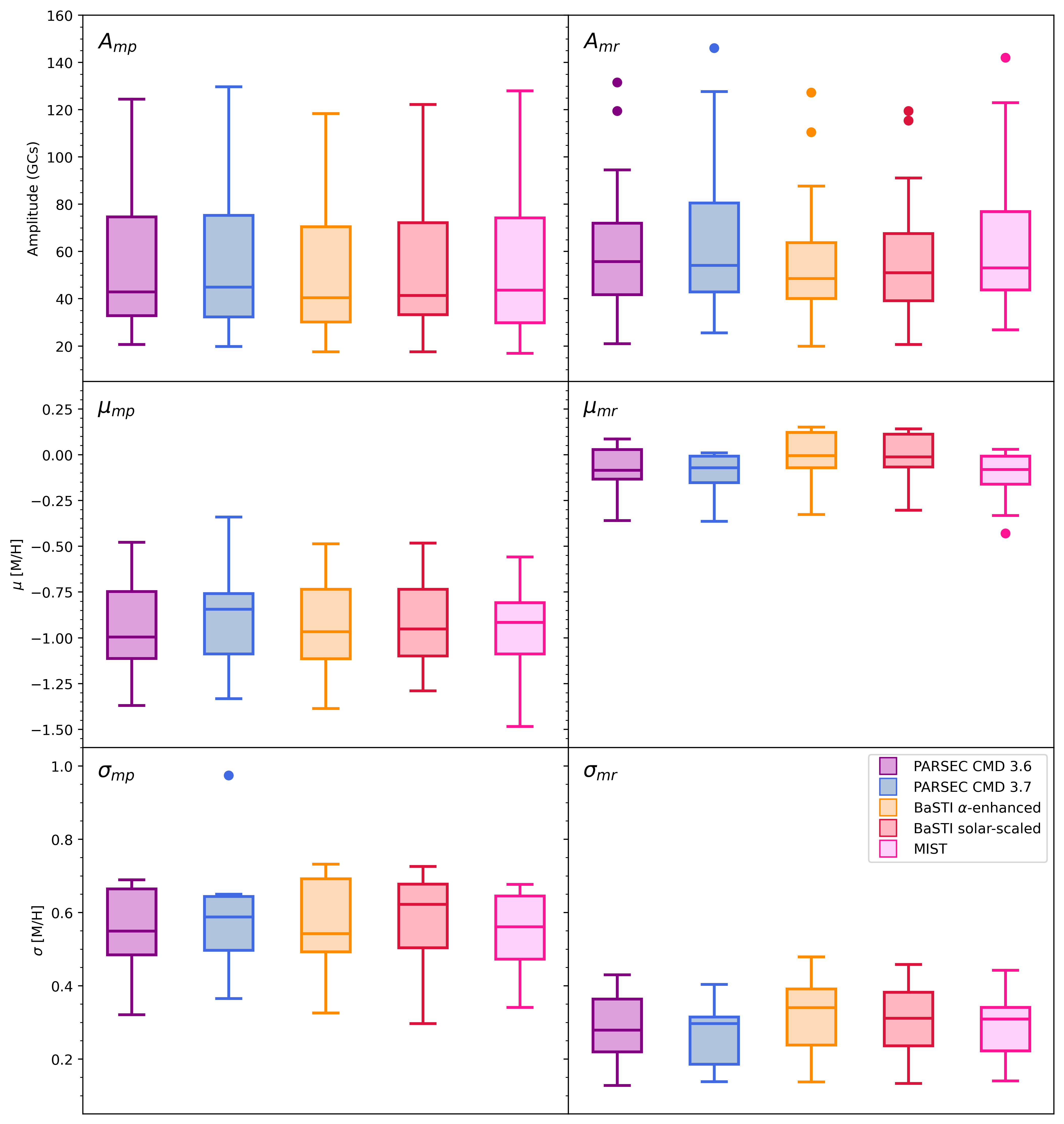}
    \caption{A comparison of double Gaussian parameters from the fitted models in Figure \ref{fig:DGfits}.  With the exception of the outlier $\sigma_{mp}$ from the PARSEC CMD 3.7 model (see the lack of a local minimum in that model for NGC 4839 in Figure \ref{fig:DGfits}) the models produce very similar sets of double Gaussian parameters.}
    \label{fig:DGbox}
\end{figure*}

Figures \ref{fig:methists} and \ref{fig:DGfits} contain the key results from this study: the shape of the MDF and the specifics of its Gaussian components are robust against the use of different stellar models and assumptions about GC age.  Both the MDFs and the fitted double Gaussian curves are visually very similar to each other for each galaxy, producing the same distribution of metallicity values within the uncertainties.  Figure \ref{fig:DGbox} displays the parameter populations in box plot form (and the double Gaussian parameters themselves can be found in Appendix \ref{app:DG_params}).

\begin{figure*}
    \centering
    \includegraphics[width=\textwidth]{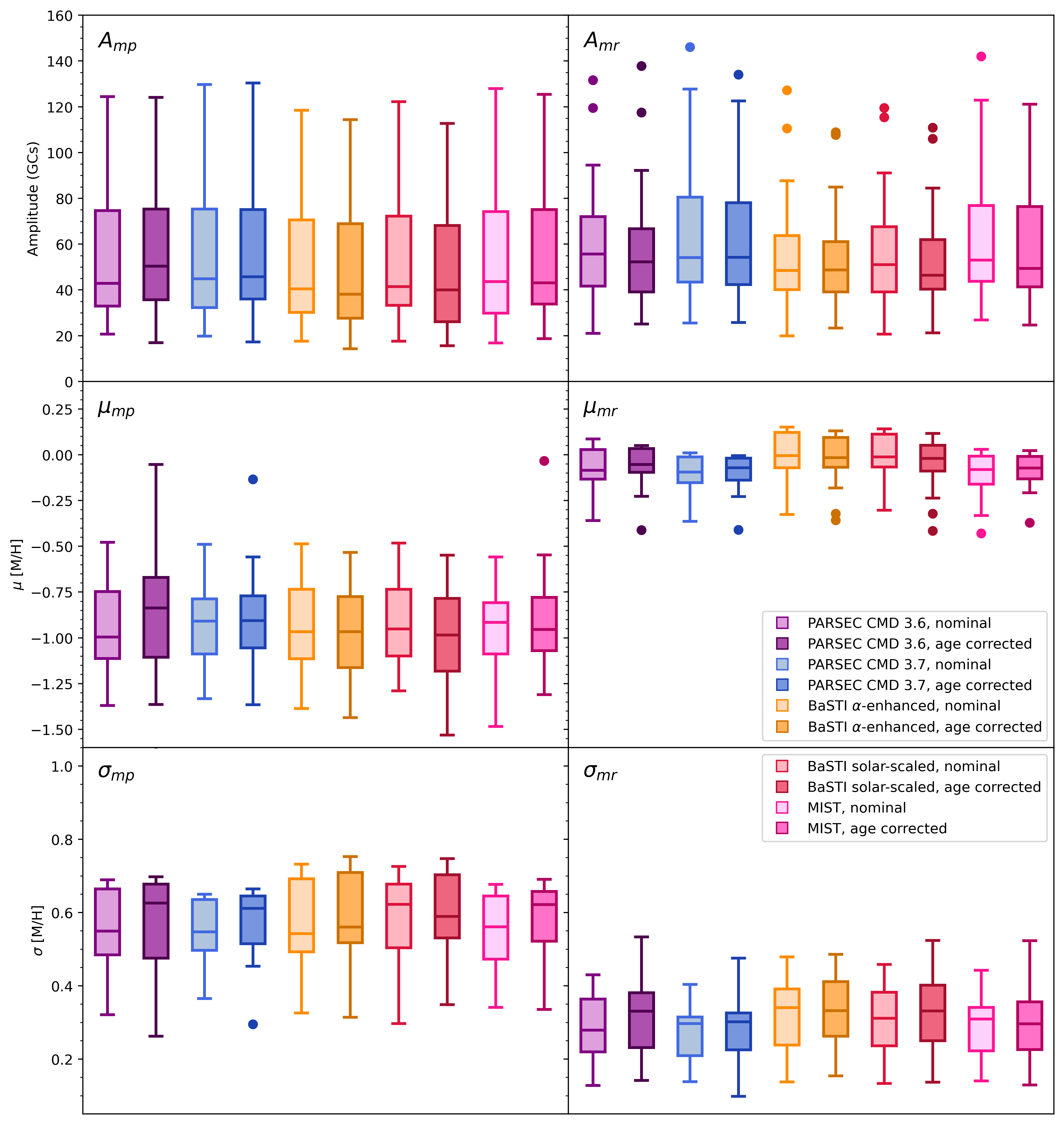}
    \caption{A comparison of double Gaussian parameters from the 5 stellar models with (lighter boxes) and without (darker boxes) age corrections.}
    \label{fig:DGboxage}
\end{figure*}

Visual inspection (see Figure \ref{fig:DGboxage}) indicates that, like stellar model choice, our built-in age correction makes no significant difference to subsequently fitted double Gaussian parameters; this was expected based on the sub-0.05-magnitude color changes that the correction produced.  The underlying reason for this lack of sensitivity appears to be simply that in this high 9- to 13-Gyr age range, the positions of the stellar red giant tracks and thus the integrated GC colors are quite insensitive to age, and it is only the cluster metallicity that is most important.  Said differently, at typical GC ages, most of the stars still present are low-luminosity and cool; beyond an age of ~8 Gyr, the position of the red giant branch in CMDs changes so slowly that age effects are secondary to metallicity effects.

Even in metallicity ranges with the most disagreement between models, the differences in derived metallicities are no greater than $\sim0.1$ dex in magnitude.  Based on expected color uncertainties and the corresponding metallicity uncertainties from Equation \ref{eq:deltaMH} and Figure \ref{fig:dmdc}, for typical globular cluster metallicities, all five models agree with each other within uncertainty.  At the GCS level, small differences in color-color transformations are effectively washed out, as demonstrated by the box plots in Figures \ref{fig:DGbox} and \ref{fig:DGboxage}.

\section{Conclusions} \label{sec:conclusions}

In this study, we have compared five suites of stellar models, using each of them in the GC color-color-metallicity conversion process from \citet{hartman2023GCsenviro} and quantifying their effects on the analysis of real GCS data.  We have also probed the effect on the CMR of adding an age-metallicity correlation drawn from the Milky Way GCs.  In all cases, regardless of stellar model choice or presence of the age correction, there was no significant change in the values used to characterize the MDFs derived from real data.

The data used in testing these models comes from 15 elliptical galaxies with very similar stellar masses and GCSs with of order $10^3$ GCs; it would be prudent to investigate whether the results hold for smaller GCSs.  In this work, we have studied the CMR primarily for the single WFC3 color index (F475X - F110W); in followup work, we will extend the transformations to other HST-based colors including the ones used in \citet{harris2023colours}.  In an upcoming study with new \textit{HST} multicolor imaging, the current heavy reliance on the stellar models will be reduced through the construction of strictly empirical transformations between color indices.

\section*{Acknowledgments}

We acknowledge financial support from the Natural Sciences and Engineering Research Council of Canada (NSERC) through a Discovery Grant to WEH.  KH thanks Laura Greggio for her advice on working with MIST, and Veronika Dornan, Claude Cournoyer-Cloutier, and Jeremy Karam for helpful discussions.

\vspace{5mm}
\facilities{\textit{HST} (WFC3)}

\software{Python (\href{https://www.python.org}{https://www.python.org}), NumPy \citep{harris2020numpy}, pandas \citep{team2020pandas}, SciPy \citep{virtanen2020scipy}, Matplotlib \citep{hunter2007matplotlib}, DOLPHOT \citep{dolphin2000dolphot}}

\appendix

\section{CMR parameters}
\label{app:relation_params}

Fitted parameters for the color-metallicity relations are listed in Table \ref{tab:nominal_fits_byage}.

\begin{table*}
	\centering
	\caption{Parameters of the fitted color CMR (see Equation \ref{eq:invexp}) for each stellar model, color index, and GC age.  These curves appear in Figure \ref{fig:metcolfits}.}
	\label{tab:nominal_fits_byage}
    \scriptsize
	\begin{tabular}{llc|ccc}
		\hline
		Index & Model & Age (Gyr) & A & B & C\\
		\hline
		(F475X - F110W) & PARSEC CMD 3.6 & 9 & 1.317 & 1.079 & 0.718\\
		  &   & 10 & 1.375 & 1.036 & 0.857\\
		  &   & 11 & 1.311 & 1.102 & 0.725\\
          &   & 12 & 1.298 & 1.133 & 0.709\\
          &   & 13 & 1.228 & 1.260 & 0.657\\
          & PARSEC CMD 3.7 & 9 & 1.177 & 1.196 & 0.564\\
          &   & 10 & 1.253 & 1.171 & 0.629\\
		  &   & 11 & 1.439 & 0.984 & 0.868\\
          &   & 12 & 1.224 & 1.240 & 0.620\\
          &   & 13 & 1.340 & 1.164 & 0.738\\
          & BaSTI $\alpha$-enhanced & 9 & -0.106 & 2.416 & 0.174\\
          &   & 10 & 0.862 & 1.456 & 0.327\\
		  &   & 11 & -4.795 & 7.103 & 0.051\\
          &   & 12 & 0.737 & 1.676 & 0.321\\
          &   & 13 & 0.652 & 1.779 & 0.299\\
          & BaSTI solar-scaled & 9 & 0.307 & 1.870 & 0.227\\
          &   & 10 & 0.758 & 1.440 & 0.310\\
		  &   & 11 & 0.888 & 1.339 & 0.336\\
          &   & 12 & 0.851 & 1.337 & 0.310\\
          &   & 13 & 0.797 & 1.538 & 0.336\\
          & MIST & 9 & 0.496 & 1.815 & 0.285\\
          &   & 10 & 0.559 & 1.781 & 0.299\\
		  &   & 11 & 0.647 & 1.722 & 0.322\\
          &   & 12 & 0.740 & 1.659 & 0.352\\
          &   & 13 & 0.870 & 1.558 & 0.399\\
          \hline
        (F475W - F850LP) & PARSEC CMD 3.6 & 9 & 1.128 & 0.911 & 0.640\\
          &   & 10 & 1.199 & 0.863 & 0.795\\
		  &   & 11 & 1.127 & 0.939 & 0.652\\
          &   & 12 & 1.120 & 0.972 & 0.652\\
          &   & 13 & 1.031 & 1.101 & 0.579\\
          & PARSEC CMD 3.7 & 9 & 0.963 & 1.054 & 0.473\\
          &   & 10 & 1.029 & 1.028 & 0.519\\
		  &   & 11 & 1.259 & 0.820 & 0.826\\
          &   & 12 & 1.003 & 1.101 & 0.525\\
          &   & 13 & 1.131 & 1.006 & 0.644\\
          & BaSTI $\alpha$-enhanced & 9 & -0.496 & 2.505 & 0.140\\
          &   & 10 & 0.614 & 1.404 & 0.282\\
		  &   & 11 & -13.91 & 15.93 & 0.019\\
          &   & 12 & 0.477 & 1.624 & 0.276\\
          &   & 13 & 0.342 & 1.777 & 0.250\\
          & BaSTI solar-scaled & 9 & 0.037 & 1.864 & 0.193\\
          &   & 10 & 0.529 & 1.392 & 0.270\\
		  &   & 11 & 0.667 & 1.281 & 0.296\\
          &   & 12 & 0.549 & 1.365 & 0.251\\
          &   & 13 & 0.554 & 1.488 & 0.290\\
          & MIST & 9 & 0.007 & 1.967 & 0.200\\
          &   & 10 & 0.088 & 1.912 & 0.212\\
		  &   & 11 & 0.200 & 1.829 & 0.231\\
          &   & 12 & 0.318 & 1.738 & 0.256\\
          &   & 13 & 0.511 & 1.571 & 0.302\\
		\hline
	\end{tabular}
\end{table*}

\section{Fitted double Gaussian parameters}
\label{app:DG_params}

Double-Gaussian fits obtained through GMM, for the giant galaxies studied in \citet{hartman2023GCsenviro}, are listed in Table \ref{tab:nominal_DG_fits}.

\begin{table*}
    \centering
    \caption{Fitted double Gaussian parameters from Equation \ref{eq:DG} for the \citet{hartman2023GCsenviro} data transformed with our nominal color-metallicity relations.  These curves appear in Figure \ref{fig:DGfits}.}
    \label{tab:nominal_DG_fits}
    \tiny
    \begin{tabular}{ll|cccccc}
        \hline
        Galaxy & Model & $A_{mp}$ (GCs) & $\mu_{mp}$ [M/H] & $\sigma_{mp}$ [M/H] & $A_{mr}$ (GCs) & $\mu_{mr}$ [M/H] & $\sigma_{mr}$ [M/H]\\
        \hline
        NGC 57 & PARSEC CMD 3.6 & 61.0 & -1.08 & 0.48 & 119.5 & -0.03 & 0.32\\
         & PARSEC CMD 3.7 & 61.0 & -1.08 & 0.47 & 127.8 & -0.07 & 0.30\\
         & BaSTI $\alpha$-enhanced & 56.2 & -1.07 & -0.52 & 110.1 & 0.06 & 0.34\\
         & BaSTI solar-scaled & 58.4 & -1.02 & 0.53 & 115.5 & 0.06 & 0.31\\
         & MIST & 59.0 & -1.10 & 0.49 & 123.0 & -0.06 & 0.31\\
         \hline
        NGC 410 & PARSEC CMD 3.6 & 40.5 & -0.85 & 0.66 & 39.4 & 0.02 & 0.21\\
         & PARSEC CMD 3.7 & 41.2 & -0.84 & 0.65 & 42.8 & -0.01 & 0.19\\
         & BaSTI $\alpha$-enhanced & 38.9 & -0.83 & 0.70 & 35.9 & 0.13 & 0.22\\
         & BaSTI solar-scaled & 39.1 & -0.83 & 0.68 & 38.6 & 0.11 & 0.22\\
         & MIST & 40.7 & -0.85 & 0.65 & 41.5 & 0.00 & 0.20\\
         \hline
        NGC 533 & PARSEC CMD 3.6 & 91.6 & -0.79 & 0.68 & 61.1 & 0.04 & 0.18\\
         & PARSEC CMD 3.7 & 94.0 & -0.79 & 0.65 & 70.9 & 0.01 & 0.17\\
         & BaSTI $\alpha$-enhanced & 84.9 & -0.78 & 0.72 & 58.7 & 0.13 & 0.20\\
         & BaSTI solar-scaled & 86.9 & -0.77 & 0.69 & 62.7 & 0.13 & 0.20\\
         & MIST & 91.8 & -0.78 & 0.67 & 63.2 & 0.02 & 0.18\\
         \hline
        NGC 741 & PARSEC CMD 3.6 & 28.0 & -1.26 & 0.32 & 41.8 & -0.34 & 0.35\\
         & PARSEC CMD 3.7 & 28.8 & -1.20 & 0.37 & 43.1 & -0.31 & 0.31\\
         & BaSTI $\alpha$-enhanced & 26.9 & -1.27 & 0.33 & 39.6 & -0.29 & 0.38\\
         & BaSTI solar-scaled & 27.6 & -1.27 & 0.30 & 40.7 & -0.30 & 0.38\\
         & MIST & 28.2 & -1.24 & 0.34 & 42.6 & -0.33 & 0.33\\
         \hline
        NGC 777 & PARSEC CMD 3.6 & 41.0 & -1.14 & 0.46 & 94.5 & -0.09 & 0.37\\
         & PARSEC CMD 3.7 & 42.1 & -1.09 & 0.49 & 99.4 & -0.09 & 0.34\\
         & BaSTI $\alpha$-enhanced & 38.6 & -1.16 & 0.48 & 87.7 & -0.01 & 0.41\\
         & BaSTI solar-scaled & 40.7 & -1.13 & 0.47 & 91.1 & -0.01 & 0.38\\
         & MIST & 41.3 & -1.08 & 0.52 & 94.8 & -0.08 & 0.35\\
         \hline
        NGC 1016 & PARSEC CMD 3.6 & 35.2 & -0.70 & 0.66 & 33.2 & 0.06 & 0.28\\
         & PARSEC CMD 3.7 & 32.9 & -0.79 & 0.63 & 39.8 & -0.01 & 0.30\\
         & BaSTI $\alpha$-enhanced & 31.4 & -0.74 & 0.69 & 31.8 & 0.12 & 0.34\\
         & BaSTI solar-scaled & 31.9 & -0.74 & 0.68 & 32.9 & 0.12 & 0.34\\
         & MIST & 31.3 & -0.83 & 0.64 & 40.4 & -0.01 & 0.31\\
         \hline
        NGC 1129 & PARSEC CMD 3.6 & 107.6 & -1.00 & 0.47 & 131.6 & -0.14 & 0.32\\
         & PARSEC CMD 3.7 & 105.8 & -1.03 & 0.44 & 146.1 & -0.20 & 0.31\\
         & BaSTI $\alpha$-enhanced & 95.5 & -1.04 & 0.48 & 127.3 & -0.11 & 0.37\\
         & BaSTI solar-scaled & 102.7 & -0.95 & 0.50 & 119.5 & -0.07 & 0.34\\
         & MIST & 102.5 & -1.05 & 0.45 & 142.0 & -0.19 & 0.32\\
         \hline
        NGC 1600 & PARSEC CMD 3.6 & 88.4 & -0.70 & 0.69 & 69.1 & -0.10 & 0.23\\
         & PARSEC CMD 3.7 & 89.8 & -0.73 & 0.64 & 81.4 & -0.11 & 0.22\\
         & BaSTI $\alpha$-enhanced & 85.7 & -0.65 & 0.73 & 63.7 & 0.00 & 0.22\\
         & BaSTI solar-scaled & 86.2 & -0.64 & 0.73 & 67.7 & -0.02 & 0.21\\
         & MIST & 89.5 & -0.71 & 0.66 & 77.5 & -0.10 & 0.21\\
         \hline
        NGC 2340 & PARSEC CMD 3.6 & 42.9 & -1.00 & 0.49 & 55.8 & -0.06 & 0.38\\
         & PARSEC CMD 3.7 & 46.6 & -0.91 & 0.53 & 56.0 & -0.04 & 0.32\\
         & BaSTI $\alpha$-enhanced & 40.5 & -0.97 & 0.54 & 49.7 & 0.05 & 0.42\\
         & BaSTI solar-scaled & 41.4 & -0.97 & 0.52 & 51.9 & 0.03 & 0.40\\
         & MIST & 45.0 & -0.92 & 0.55 & 51.3 & -0.03 & 0.35\\
         \hline
        NGC 3158 & PARSEC CMD 3.6 & 26.6 & -1.24 & 0.49 & 41.6 & -0.17 & 0.43\\
         & PARSEC CMD 3.7 & 28.2 & -1.17 & 0.52 & 43.9 & -0.17 & 0.37\\
         & BaSTI $\alpha$-enhanced & 25.4 & -1.26 & 0.47 & 40.6 & -0.12 & 0.46\\
         & BaSTI solar-scaled & 25.0 & -1.29 & 0.47 & 41.9 & -0.14 & 0.46\\
         & MIST & 26.0 & -1.28 & 0.46 & 44.9 & -0.21 & 0.42\\
         \hline
        NGC 3842 & PARSEC CMD 3.6 & 30.6 & -1.37 & 0.52 & 52.2 & -0.36 & 0.43\\
         & PARSEC CMD 3.7 & 31.8 & -1.33 & 0.50 & 54.2 & -0.36 & 0.40\\
         & BaSTI $\alpha$-enhanced & 29.0 & -1.39 & 0.51 & 48.6 & -0.33 & 0.48\\
         & BaSTI solar-scaled & 34.7 & -1.14 & 0.62 & 39.4 & -0.21 & 0.42\\
         & MIST & 28.5 & -1.48 & 0.42 & 58.3 & -0.43 & 0.44\\
         \hline
        NGC 4073 & PARSEC CMD 3.6 & 43.5 & -0.84 & 0.69 & 74.9 & -0.12 & 0.28\\
         & PARSEC CMD 3.7 & 44.9 & -0.84 & 0.65 & 79.8 & -0.14 & 0.27\\
         & BaSTI $\alpha$-enhanced & 45.6 & -0.73 & 0.73 & 63.9 & -0.01 & 0.27\\
         & BaSTI solar-scaled & 46.0 & -0.73 & 0.71 & 67.7 & -0.01 & 0.26\\
         & MIST & 43.6 & -0.84 & 0.68 & 76.4 & -0.13 & 0.27\\
         \hline
        NGC 4839 & PARSEC CMD 3.6 & 44.6 & -1.02 & 0.55 & 45.3 & -0.09 & 0.27\\
         & PARSEC CMD 3.7 & 45.0 & -1.00 & 0.55 & 46.9 & -0.12 & 0.25\\
         & BaSTI $\alpha$-enhanced & 42.0 & -1.05 & 0.54 & 46.1 & -0.03 & 0.29\\
         & BaSTI solar-scaled & 42.2 & -1.07 & 0.51 & 51.0 & -0.06 & 0.29\\
         & MIST & 43.8 & -1.01 & 0.56 & 45.8 & -0.11 & 0.26\\
         \hline
        NGC 4914 & PARSEC CMD 3.6 & 20.7 & -0.48 & 0.57 & 21.2 & 0.09 & 0.20\\
         & PARSEC CMD 3.7 & 19.9 & -0.49 & 0.59 & 25.6 & 0.01 & 0.18\\
         & BaSTI $\alpha$-enhanced & 17.6 & -0.49 & 0.64 & 20.0 & 0.15 & 0.26\\
         & BaSTI solar-scaled & 17.6 & -0.48 & 0.63 & 20.7 & 0.14 & 0.25\\
         & MIST & 16.9 & -0.62 & 0.60 & 26.9 & 0.01 & 0.23\\
         \hline
        NGC 7242 & PARSEC CMD 3.6 & 124.5 & -0.55 & 0.62 & 58.7 & 0.05 & 0.13\\
         & PARSEC CMD 3.7 & 129.8 & -0.56 & 0.60 & 53.4 & -0.57 & 0.14\\
         & BaSTI $\alpha$-enhanced & 118.5 & -0.50 & 0.67 & 41.7 & 0.12 & 0.14\\
         & BaSTI solar-scaled & 122.2 & -0.48 & 0.66 & 38.9 & 0.12 & 0.13\\
         & MIST & 128.1 & -0.56 & 0.60 & 53.1 & 0.03 & 0.14\\
         \hline
    \end{tabular}
\end{table*}

\bibliography{Bibliography}{}
\bibliographystyle{aasjournal}

\end{document}